\documentclass[11pt]{article}
\usepackage{graphicx}
\usepackage{cite}

\begin{document}

%
% Title, Authors, Affiliations
% ===================
%
\title{\Large\bf Transverse single spin asymmetry in the Drell-Yan process}

\author{Jian Zhou, Andreas Metz
 \\[0.3cm]
{\normalsize\it Department of Physics, Barton Hall,} %\\
{\normalsize\it Temple University, Philadelphia, PA 19122-6082, USA}}

\maketitle

% headline
%\thispagestyle{fancy}
%\fancyhead[R]{\tt \jobname}

%
% Abstract
% ======
%
\begin{abstract}
\noindent
We revisit the transverse single spin asymmetry in the angular distribution of a
Drell-Yan dilepton pair.
We study this asymmetry by using twist-3 collinear factorization, and we obtain the 
same result both in covariant gauge and in the light-cone gauge.
Moreover, we have checked the electromagnetic gauge invariance of our calculation.
Our final expression for the asymmetry differs from all the previous results given 
in the literature.
The overall sign of this asymmetry is as important as the sign of the Sivers 
asymmetry in Drell-Yan.
\end{abstract}

%
% 1. Section: Introduction
% ==================
%
\section{Introduction}
\noindent
The observation of transverse single spin asymmetries (SSAs) in various hard scattering 
processes has stimulated new remarkable developments both on the theoretical and the
experimental side.
As a consequence, the study of SSAs currently represents a very active field 
of research~\cite{D'Alesio:2007jt,Burkardt:2008jw,Barone:2010ef}.
The interest in such effects is essentially twofold: first, SSAs allow one to address 
the parton structure of the nucleon beyond the collinear parton model approximation.
Second, SSAs are ideal observables in order to further explore in which cases the 
machinery of QCD factorization still applies and in which cases, in its simplest form, 
it breaks down (see~\cite{Rogers:2010dm} and references therein).

For what concerns the parton structure of the nucleon, in the present work we 
focus on collinear twist-3 quark-gluon-quark correlations.
To be more precise, the central non-perturbative correlator is the so-called ETQS 
(Efremov-Teryaev-Qiu-Sterman) matrix 
element~\cite{Efremov:1981sh, Efremov:1984ip,Qiu:1991pp} $T_F$ --- and its chiral-odd 
partner $T_F^{(\sigma)}$ --- which typically appears when describing transverse SSAs 
in the context of collinear higher-twist factorization.
The machinery of collinear twist-3 factorization was pioneered already in the early
1980's~\cite{Efremov:1981sh,Ellis:1982wd,Jaffe:1991kp},
and in the meantime frequently applied to transverse spin effects in hard semi-inclusive 
reactions (see, e.g., 
Refs.~\cite{Efremov:1984ip,Qiu:1991pp,Kouvaris:2006zy,Eguchi:2006mc,Zhou:2009jm}).

In this paper, we revisit the transverse single spin asymmetry in the angular distribution 
of a Drell-Yan dilepton pair.
This asymmetry is defined as the difference of two spin dependent cross sections 
with opposite directions of transverse polarization divided by their sum,
\begin{equation} \label{eq:AN}
A_N = \bigg( \frac{d \sigma(S_T)}{d \Omega d Q^2} - \frac{d \sigma(-S_T)}{d \Omega d Q^2} \bigg)
\bigg/ 
\bigg( \frac{d \sigma(S_T)}{d \Omega d Q^2} + \frac{d \sigma(-S_T)}{d \Omega d Q^2} \bigg) \,,
\end{equation}
where $d\Omega = d \cos \theta d \phi_S$ is a solid angle element of the leptons in a
dilepton rest frame, and the azimuthal angle $\phi_S$ is measured relative to the
transverse spin vector.
Note that the transverse momentum $Q_T$ of the dilepton pair is integrated out, and we 
emphasize that integrating over $Q_T$ is essential for applying the collinear factorization
approach in the present case. 

The asymmetry $A_N$ was already studied in several previous articles, and various different 
results were obtained.
The first calculation, carried out in the light-cone gauge, can be found in 
Ref.~\cite{Hammon:1996pw}. 
The authors obtained$\,$\footnote{To shorten the notation we suppress throughout terms where 
quarks and antiquarks are interchanged.} 
\begin{equation} \label{eq:HTS}
A_N^{(HTS)} = - \, \frac{1}{Q} \, \frac{\sin 2\theta \sin \phi_S}{1+ \cos^2 \theta} \,
\frac{\sum_q e_q^2 \int dx \, \Big( T_F^q(x,x) - x \frac{d}{dx} T_F^q(x,x) \Big) f_1^{\bar{q}}(x')}
     {\sum_q e_q^2 \int dx \, f_1^q(x) f_1^{\bar{q}}(x')} \,,
\end{equation}
where $f_1^q$ is the standard unpolarized twist-2 parton distribution for quark flavor $q$.
The momentum fraction $x'$ is given by $x' = Q^2/(x S)$, with $S = (P + \bar{P})^2$ 
denoting the square of the {\it cm} energy of the process.
Later on the presence of the derivative term in the numerator of~(\ref{eq:HTS}) was doubted,
and it was argued that the correct result for $A_N$ should be~\cite{Boer:1997bw,Boer:1999si}
\begin{equation} \label{eq:BMT}
A_N^{(BMT)} = - \, \frac{1}{Q} \, \frac{\sin 2\theta \sin \phi_S}{1+ \cos^2 \theta} \,
\frac{\sum_q e_q^2 \int dx \, T_F^q(x,x) f_1^{\bar{q}}(x')}
     {\sum_q e_q^2 \int dx \, f_1^q(x) f_1^{\bar{q}}(x')} \,.
\end{equation}
Afterwards, in Ref.~\cite{Boer:2001tx} $A_N$ was considered in the collinear twist-3 
approach, and the result of that study agreed with the expression in~(\ref{eq:BMT}).
Then $A_N$ was computed by using factorization in terms of transverse momentum 
dependent correlators~\cite{Ma:2003ut}.
The final outcome of that work neither agreed with~(\ref{eq:HTS}) nor 
with~(\ref{eq:BMT}).
More recently, $A_N$ was again considered in Ref.~\cite{Anikin:2010wz}, where the
authors claimed that the spin-dependent hadronic tensor should be multiplied by a factor 
of 2 compared to previous work~\cite{Hammon:1996pw,Boer:1997bw,Boer:1999si,Boer:2001tx}. 
The controversy about the derivative term was not addressed in~\cite{Anikin:2010wz}.

This somewhat unclear situation motivated us to revisit this topic.
We computed $A_N$ in~(\ref{eq:AN}) by means of twist-3 collinear factorization and came 
up with yet another result.
To be specific, our result is just half of the one quoted in Eq.~(\ref{eq:BMT}).
An important difference in comparison to previous work is that, when performing the
collinear expansion in the twist-3 formalism, we take into account the dependence 
on transverse parton motion ($k_T$-dependence) not only in the hadronic 
tensor but also in the lepton tensor.
In order to gain further confidence we checked our calculation in a few different 
ways.

The rest of the paper is organized as the follows.
In the next section, we introduce our notation, and give some details about the kinematics.
In Section~3, we derive the asymmetry in a covariant gauge as well as in the light cone gauge, 
and the two results agree with each other.
In addition, we have checked the electromagnetic gauge invariance by explicit calculation.
We summarize the paper in Section~4.

%
% 2. Section: Kinematics and notation
% ==================
%
\section{Kinematics and notation}
We focus on lepton pair production in hadronic scattering which comes from the decay of a 
virtual photon, $H_a + H_b \to \gamma^*+X \to \ell^+ + \ell^-+X$.
The 4-momenta of the leptons are $l_1$ and $l_2$, and $q = l_1 + l_2$ denotes the
momentum of the virtual photon.
The invariant mass of the dilepton pair is $Q$ with $Q^2 = q^2$.
For the following calculation we need to introduce the vector $R = l_1 - l_2$.
In any dilepton rest frame, $R$ reads
\begin{equation} \label{eq:R}
R = Q \, \Big( 0, \, \sin \theta \cos \phi, \, \sin \theta \sin \phi, \,  \cos \theta \Big) \,, 
\end{equation}
where the numerical values of $\theta$ and $\phi$ depend on the frame.
The correlation associated with $A_N$ in~(\ref{eq:AN}) is 
$\varepsilon_{\mu\nu\rho\sigma} P^{\mu} \bar{P}^{\nu} S^{\rho} R^{\sigma}$, while
the asymmetry usually associated with the Sivers effect~\cite{Sivers:1989cc} is related 
to the correlation $\varepsilon_{\mu\nu\rho\sigma} P^{\mu} \bar{P}^{\nu} S^{\rho} q^{\sigma}$.
The latter requires to measure the transverse momentum $Q_T$ of the dilepton pair.

A convenient way of sorting out the different angular dependences of the Drell-Yan cross 
section is to decompose the lepton tensor in terms of individual independent orthogonal 
tensors~\cite{Meng:1995yn,Boer:2006eq,Berger:2007si},
\begin{equation} \label{eq:dec_1}
L^{\mu \nu}= \Big( (q+R)^\mu (q-R)^\nu + (q+R)^ \nu(q-R)^\mu - 2Q^2 g^{\mu\nu} \Big)
= \sum_i^9  L_i V^{\mu \nu}_i \,.
\end{equation}
(A discussion of the general structure of the polarized Drell-Yan cross section can
be found in Ref.~\cite{Arnold:2008kf}.)
The $L_i$ represent the angular structures, and the basis tensors $V^{\mu \nu}_i$ can
be constructed from a set of (4-dimensional) basis vectors $T^{\mu}$, $X^{\mu}$,
$Y^{\mu}$, $Z^{\mu}$, which are mutually orthogonal to each other and are normalized
according to $T^2 = 1$, $X^2 = Y^2 = Z^2 = -1$.
For the case of $A_N$, the relevant angular structures appear in the terms associated 
with $V_3 = -\frac{1}{2} (Z^\mu X^\nu + Z^\nu X^\mu)$ and 
$V_8 = -\frac{1}{2} ( Z^\mu Y^\nu + Z^\nu Y^\mu )$,
\begin{equation} \label{eq:dec_2}
L^{\mu \nu}= Q^2 \sin 2 \theta  \cos \phi  \ V^{\mu \nu}_3 +
Q^2 \sin 2 \theta  \sin \phi  \ V^{\mu \nu}_8 + \ ... \; .
\end{equation}
Like in the case of~(\ref{eq:R}), the decomposition~(\ref{eq:dec_2}) holds in any dilepton 
rest frame.

For $Q_T = 0$ (in the hadronic {\it cm} frame), we choose the following dilepton rest frame:
$z$-axis along the direction of the polarized hadron, and $x$-axis along the direction 
of the polarization vector $S_T$. 
To be fully specific, the 4-dimensional basis vectors are given by
\begin{eqnarray} \label{eq:basis_CM}
T^\mu & = & \frac{q^\mu}{\sqrt{Q^2}} \,,
\nonumber \\
Z^\mu & = & \frac{1}{Q} \, \Big( x P^{\mu} - x' \bar{P}^{\mu} \Big) \,,
\nonumber \\
X^\mu & = & S^\mu_T \,, \phantom{\frac{1}{1}}
\nonumber \\
Y^\mu & = & \varepsilon^{\mu \nu \rho \sigma } T_\nu Z_\rho X_\sigma \,. \phantom{\frac{1}{1}}
\end{eqnarray}
Because of the specific definition of $Z^{\mu}$, this frame can actually be considered
as partonic {\it cm} frame.

If $Q_T \neq 0$, one may work in the Collins-Soper frame~\cite{Collins:1977iv} for which
the basis vectors read
\begin{eqnarray} \label{eq:basis_CS}
T^\mu & = & \frac{q^\mu}{\sqrt{Q^2}} \,,
\nonumber \\
Z^\mu & = & \frac{2}{\sqrt{Q^2+Q_\perp^2}} 
\Big( q_{\bar{p}} \tilde{P}^\mu - q_p \tilde{\bar{P}}^\mu \Big) \,,
\nonumber \\
X^\mu & = & - \frac{Q}{Q_\perp}  \frac{2}{\sqrt{Q^2+Q_\perp^2}}
\Big( q_{\bar{p}} \tilde{P}^\mu + q_p \tilde{\bar{P}}^\mu \Big) \,,
\nonumber \\
Y^\mu & = & \varepsilon^{\mu \nu \rho \sigma } T_\nu Z_\rho X_\sigma \,. \phantom{\frac{1}{1}}
\end{eqnarray}
In~(\ref{eq:basis_CS}) we use the further definitions 
$\tilde{P}^\mu = [P^\mu-(P \cdot q)/q^2 q^\mu ]/ \sqrt{S}$,
$\tilde{\bar{P}}^\mu = [\bar{P}^\mu-(\bar{P} \cdot q)/q^2 q^\mu ]/ \sqrt{S}$,
with $q_p = P \cdot q /\sqrt{S}$, $q_{\bar{p}} = \bar{P} \cdot q / \sqrt{S}$.
At tree level, $Q_T$ is equal to the sum of the intrinsic transverse momenta
of the two incoming partons.
This implies that for $Q_T \neq 0$ a $k_T$-dependence is sitting in the unit 
vectors $X^\mu$, $Y^\mu$ and $T^\mu$.
(The $k_T$-dependence of $Z^\mu$ is of the order $k_T^2$ and therefore irrelevant for
our twist-3 calculation.)
As a result, the terms containing $\cos \phi$ and $\sin \phi$ are $k_T$-dependent.
This $k_T$-dependence must be taken into account when performing the collinear 
expansion.

For $Q_T \neq 0$, instead of using the Collins-Soper frame, one can alternatively perform 
the calculation, for instance, in the Gottfried-Jackson frame\cite{Gottfried:1964nx}.  
Keeping track of all $k_T$-dependent terms in the Gottfried-Jackson frame is more
involved.
Nevertheless, we carried out the calculation, and our final result agrees with what we
find in the Collins-Soper frame.

%
% 3. Section: Calculation in twist-3 collinear factorization
% ==================
%
\section{Calculation in twist-3 collinear factorization}
In order to calculate $A_N$ in Eq.~(\ref{eq:AN}) one needs both the unpolarized
cross section (in the parton model) and the spin-dependent cross section.
The former is well-known and given by
\begin{equation}
\frac{d \sigma}{d Q^2 d\Omega} = \frac{4 \pi \alpha_{em}^2}{9 Q^2}
\sum_{q} e_q^2 \int dx \, dx' \, f_1^{q}(x) \, f_1^{\bar{q}}(x') \,
\bigg[ \frac{3}{16\pi} \, (1+\cos^2 \theta) \, 
\delta \Big( Q^2 -x x' S \Big) \bigg] \,.
\end{equation}
The polarized cross section is a twist-3 effect and depends on quark-gluon-quark 
correlations, which contain interesting physics beyond the parton model.
In fact, such twist-3 correlations associated with both hadrons can give rise to  
$A_N$ leading to the generic expression~\cite{Boer:1997bw}
\begin{equation}
A_N \propto \frac{1}{Q} \,
\frac{ T_F(x,x) \otimes f_1(x') + 
       h_1(x) \otimes T_F^{(\sigma)}(x',x') }
     { f_1(x) \otimes f_1(x')} \,,
\end{equation}
where $h_1$ is the transversity distribution. 
The second (chiral-odd) term in the numerator, which we have not included in 
Eqs.~(\ref{eq:HTS}) and~(\ref{eq:BMT}), was first considered in Ref.~\cite{Boer:1997bw}.
In our calculation we treat both the chiral-even and the chiral-odd contribution 
to $A_N$.

The ETQS matrix element $T_F$ and its chiral-odd partner $T_F^{(\sigma)}$ are
defined as$\,$\footnote{For a generic 4-vector $v$, we define light-cone coordinates
according to $v^{\pm} = (v^0 \pm v^3) / \sqrt{2}$ and $\vec{v}_T = (v^1,v^2)$.}
\begin{eqnarray}
T_F(x,x_1) & = & \int \frac{dy^- dy^-_1}{4 \pi} \, e^{-ixP^+ y^-+i(x_1-x)P^+y_1^-} 
\langle PS | \bar{\psi}(y^-) \, \gamma^{+} \, \varepsilon_T^{\nu\mu} \, S_{T \nu} \,
g F_{\phantom{+} \mu}^{+}(y_1^-) \, \psi(0) | PS \rangle \,,
\nonumber \\
T_F^{(\sigma)}(x,x_1) & = & \int \frac{dy^- dy^-_1}{ 4 \pi} \, e^{-ixP^+ y^-+i(x_1-x)P^+y_1^-} 
\langle PS | \bar{\psi}(y^-) \, \sigma^{\mu +} \,
g F_{\phantom{+} \mu}^{+}(y_1^-) \, \psi(0) | PS \rangle \,,
\end{eqnarray}
where a summation over color is implicit, and gauge links has been suppressed.
In the following two subsections we compute the hard coefficients associated with these 
matrix elements both in covariant gauge and in the light-cone gauge.
It is worthwhile to mention that $T_F(x,x)$ and $T_F^{(\sigma)}(x,x)$ are related to
particular $k_T$-moments of the transverse momentum dependent Sivers 
function~\cite{Sivers:1989cc} and Boer-Mulders function~\cite{Boer:1997nt}, 
respectively~\cite{Boer:2003cm,Ma:2003ut}.

%
% 3.1 Subsection: Asymmetry derived in covariant gauge}
% ==================
%
\subsection{Asymmetry derived in covariant gauge}
\begin{figure}[t]
\begin{center}
\includegraphics[width=12cm]{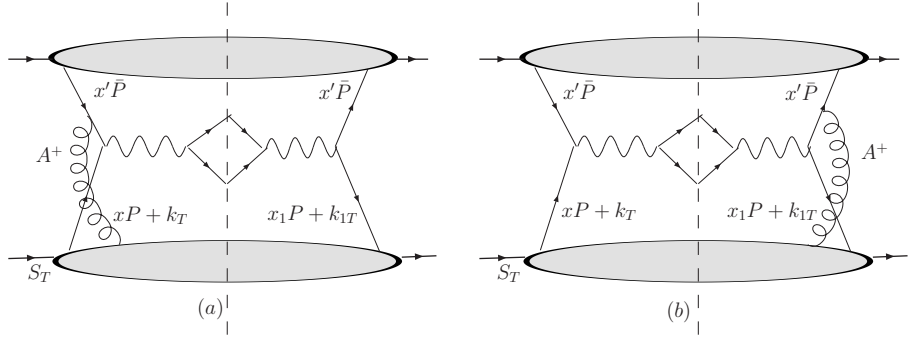}
\end{center}
\vskip -0.4cm
\caption{Diagrams contributing to $A_N$ in covariant gauge.
The gluon attached to the hard scattering part is longitudinally polarized.
In order to extract the twist-3 contribution, one has to expand in $k_T$ and 
$k_{1T}$, and to pick up the linear terms.} \label{f:1}
\end{figure}

In covariant gauge, the leading contribution of the gluon field is from the component parallel 
to the direction of its momentum.
If one considers $P^+$ (with $P$ being the momentum of the polarized nucleon) and 
$\bar{P}^-$ as the large light-cone momenta, then the dominant component of the gluon
field for the diagrams shown in Fig.~\ref{f:1} is $A^{+}$.
Before making the collinear expansion, the incoming partons carry a transverse momentum
$k_{iT}$, which is much smaller than the dominant longitudinal momentum.
In order to extract the twist-3 contributions from the diagrams with one-gluon-exchange,
one needs to get one power of $k_{iT}$ from the hard scattering part and combine $k_{iT}$ 
with $A^+$ in order to convert the gluon field in the matrix element into the corresponding 
part of the field strength tensor ($\partial_T A^+$)~\cite{Qiu:1991pp}.
As stated above, the $k_{T}$-flow may go through the lepton lines via the virtual photon. 
%since the transverse momentum of the dilepton pair has been integrated out.
Therefore, we have to expand the hadronic tensor as well as the lepton tensor in terms of 
$k_{iT}$ around $k_{iT} = 0$.
%%%%%%%%%%%%%% Figure 1 %%%%%%%%%%%%%%

Only the two diagrams in Fig.~\ref{f:1} contribute to $A_N$ in covariant gauge.
To be more precise, these two diagrams provide the chiral-even $T_F$ part of the 
asymmetry.
In order to get the chiral-odd contribution one has to consider the corresponding 
two diagrams for which the gluon is associated with the unpolarized hadron.
As an example, for the left cut-diagram in Fig.~\ref{f:1} we have the following 
expansion,
\begin{eqnarray} \label{eq:cov_detail}
\lefteqn{
H^{\mu \nu, \rho}(xp+k_T,x_1p+k_{1T},S_T) \, P_\rho \, L_{\mu \nu}(q=x_1p+k_{1T}+x' \bar p,R)}
\nonumber \\
& = &
H^{\mu \nu, \rho}(xp,x_1p,S_T) \, P_\rho \, L_{\mu \nu}(q=x_1p+x' \bar p,R) 
\phantom{\frac{1}{1}}
\nonumber \\
& & + Q^2  \sin 2 \theta
 \Bigg[ \frac{\partial \Big( \cos \phi \, V^{CS}_{3,\mu \nu} \,
  H^{\mu \nu,\rho} \, P_\rho \Big)}{\partial k_{1T}^\sigma} +
        \frac{\partial \Big( \sin \phi \, V^{CS}_{8,\mu \nu} \,
  H^{\mu \nu,\rho} \, P_\rho \Big)}{\partial k_{1T}^\sigma} 
\Bigg]_{k_T = k_{1T} = 0}  k_{1T}^{\sigma}
\nonumber \\
& & + Q^2  \sin 2 \theta
 \Bigg[ \sin \phi_{S} \, V^{CM}_{8,\mu \nu} 
 \frac{\partial  H^{\mu \nu,\rho} \, P_\rho}{\partial k_{T}^\sigma} 
\Bigg]_{k_T = k_{1T} = 0}  k_{T}^{\sigma}
 + ... \,,
\end{eqnarray}
where the superscripts $CM$ and $CS$ refer to the partonic {\it cm} frame 
and the Collins-Soper frame specified in~(\ref{eq:basis_CM}) and in~(\ref{eq:basis_CS}), 
respectively.
The azimuthal angle $\phi$ is understood in the Collins-Soper frame, while the
azimuthal angle in the {\it cm} frame is just what we defined above as $\phi_S$, 
namely the angle between $R_T$ and $S_T$.
There is no need to distinguish between the polar angle $\theta$ in the two frames
when expanding around $k_{iT} = 0$ and keeping only the linear terms.
For the left cut-diagram in Fig.~\ref{f:1}, the lepton tensor is independent of $k_{T}$,
but it depends on $k_{1T}$.
The used tensor decomposition of the lepton tensor is rather convenient in order to 
treat this $k_{1T}$-dependence. 
This dependence is sitting in three parts: the angular dependences $\cos \phi$ and 
$\sin \phi$, the tensors $V^{CS}_{3,\mu \nu}$ and $V^{CS}_{8,\mu \nu}$, and
the hadronic tensor $H^{\mu \nu,\rho} \, P_\rho$.

The first term of the Taylor expansion in~(\ref{eq:cov_detail}) corresponds to the eikonal 
line contribution to the twist-2 quark distribution, which does not contribute to the 
asymmetry.
One can extract the desired twist-3 term by picking up the terms linear in $k_T$ 
(and $k_{1T}$) from the above expansion.
Note that in $H^{\mu \nu,\rho} \, P_\rho$ also a delta function of the form
$\delta (Q^2 - (xp+k_{1T} +x'\bar{p})^2 )$ is hidden.
It is easy to see that this delta-function cannot provide a term linear in $k_{1T}$,
and therefore its $k_{1T}$-dependence is irrelevant for the calculation of $A_N$.
This is actually the reason why the derivative term of $T_F$, which we briefly 
discussed in the Introduction, does not show up in $A_N$.
In general, the collinear expansion enables one to integrate out three of the four 
components of the parton loop momenta, and as a result the non-perturbative part can 
be expressed through the collinear twist-3 correlations $T_F$ and $T_F^{(\sigma)}$.

The strong interaction phase necessary for having a nonzero SSA arises from the partonic
scattering amplitude with an extra gluon. 
As is evident from the diagrams in Fig.~\ref{f:1}, this amplitude interferes with the 
real scattering amplitude without a gluon.
The imaginary part is due to the pole of the quark (antiquark) propagator and arises 
when integrating over the longitudinal gluon momentum fraction $x_g$.
In the present case, one has a pole for $x_g = 0$ (``soft gluon pole'' from initial
state interaction), while there is no contribution from so-called hard gluon poles or 
soft fermion poles.
We extract the imaginary part of the pole by using the formula
\begin{equation}
{\rm Im} \, \frac{1}{x_g \pm i\epsilon}= \mp \, i \pi \delta(x_g).
\end{equation}
Collecting all the pieces we finally arrive at the following polarized differential cross 
section,
\begin{eqnarray}
\frac{d \sigma(S_T)}{d Q^2 d\Omega} & = & \frac{4 \pi\alpha_{em}^2}{9 Q^2} 
\sum_{q} e_q^2 \int dx \, dx' \, 
\Big( T_F^q(x,x) \, f_1^{\bar{q}}(x') + h_1^q(x) \, T_{F}^{(\sigma) \, \bar{q}}(x',x') \Big)
\nonumber \\ 
& & \hspace{3.0cm} \mbox{} \times
\frac{1}{Q} \bigg[ \frac{3}{32\pi} \, (- \sin 2 \theta \sin \phi) \,
 \delta \Big( Q^2-xx' S \Big) \bigg] \,.
\end{eqnarray}
This provides the asymmetry
\begin{equation} \label{eq:final}
A_N = - \, \frac{1}{2Q} \, \frac{\sin 2\theta \sin \phi_S}{1+ \cos^2 \theta} \,
\frac{\sum_q e_q^2 \int dx \, 
\Big( T_F^q(x,x) \, f_1^{\bar{q}}(x') + h_1^q(x) \, T_{F}^{(\sigma) \, \bar{q}}(x',x') \Big)}
     {\sum_q e_q^2 \int dx \, f_1^q(x) f_1^{\bar{q}}(x')} \,,
\end{equation}
which, as already stated above, is just half of the result~(\ref{eq:BMT}) obtained in 
Refs.~\cite{Boer:1997bw,Boer:1999si,Boer:2001tx}.

%
% 3.2 Subsection: Asymmetry derived in the light-cone gauge}
% ==================
%
\subsection{Asymmetry derived in the light-cone gauge}
\begin{figure}[t]
\begin{center}
\includegraphics[width=6cm]{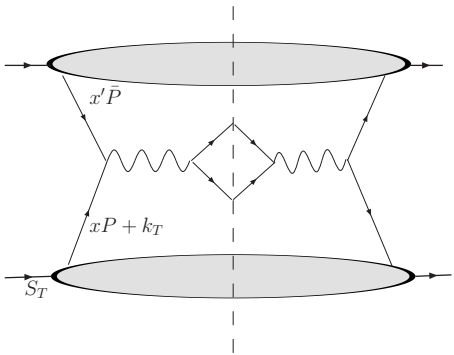}
\end{center}
\vskip -0.4cm 
\caption{Contribution from the $k_T$-expansion in the light cone gauge.
The $k_T$-flow goes also through the lepton lines via the virtual 
photon propagator.} \label{f:2}
\end{figure}

\begin{figure}[t]
\begin{center}
\includegraphics[width=12cm]{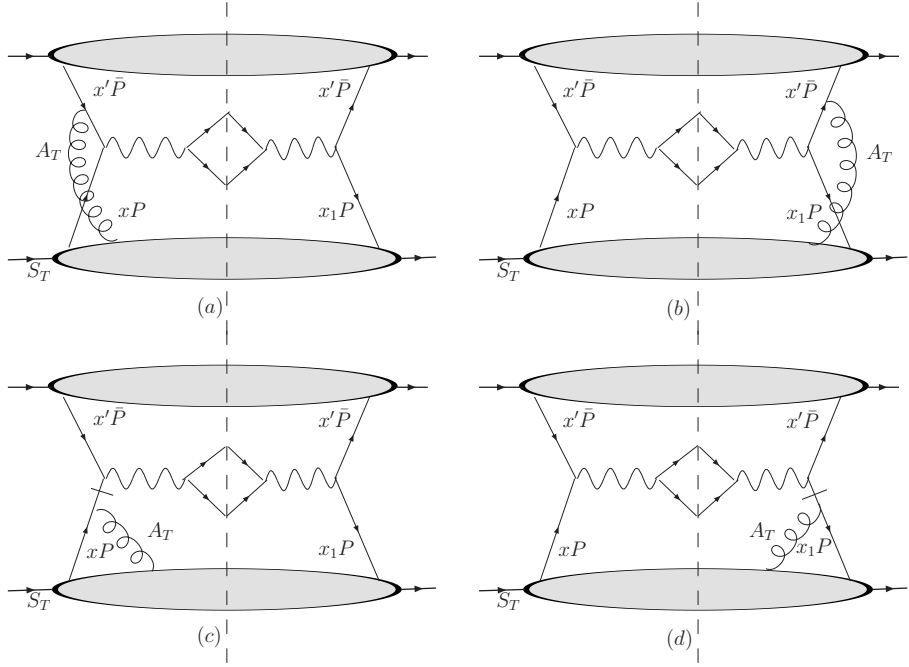}
\end{center}
\vskip -0.4cm 
\caption{Feynman diagrams with one-gluon-exchange relevant for the calculation 
of $A_N$ in the light-cone gauge.
The momenta carried by all incoming partons have only a longitudinal component.
The diagrams~(c) and~(d) represent the contribution from the so-called special 
fermion propagator introduced in Ref.~\cite{Qiu:1988dn}.
Note that the special propagator actually contributes to the hard coefficients
of both the operator $\bar{\psi} \partial_T \psi$ and the operator 
$\bar{\psi} A_T \psi$.
One finds that these two contributions exactly cancel each other.} \label{f:3}
\end{figure}

To test the color gauge invariance of our result, we derived the asymmetry also in 
the color light-cone gauge.
In general, in the light-cone gauge both the first order $k_T$-expansion of the born 
diagram (see Fig.~\ref{f:2}) and the diagrams with one additional exchange of a 
transversely polarized gluon (see Fig.~\ref{f:3}) contribute to the spin dependent 
cross section at the twist-3 level.
The associated twist-3 non-perturbative parts are the matrix elements for which the 
operators $\bar{\psi} \partial_T \psi$ and $\bar{\psi} A_T \psi$ are sandwiched between 
the hadron state~\cite{Ellis:1982wd}.
Apparently, these two correlators are not QCD gauge invariant.
However, if one entirely fixes the light-cone gauge, i.e., if one carries out the 
calculation using a specific boundary condition for the transverse gluon field at the 
light-cone infinity, then the two matrix elements can be uniquely related to the gauge
invariant quark-gluon-quark correlators $T_F$ and 
$T_F^{(\sigma)}$~\cite{Zhou:2009jm,Zhou:2008mz}.
%%%%%%%%%%%%%% Figure 2 %%%%%%%%%%%%%%
%%%%%%%%%%%%%% Figure 3 %%%%%%%%%%%%%%

There exist three frequently used boundary conditions: the retarded boundary 
condition, the advanced boundary condition, and the anti-symmetric boundary condition.
For the Drell-Yan process the retarded boundary condition $A_T(- \infty^-) = 0$ is
the most convenient choice~\cite{Zhou:2009rp}. 
Exploiting this particular boundary condition, the operators 
$\bar{\psi} \partial_T \psi$ and $\bar{\psi} A_T \psi$ can be readily rewritten in
a gauge invariant form.
For example, one has 
\begin{equation} \label{eq:lc_1}
\int \frac{dy^-}{4 \pi} \, e^{ixP^+ y^-} \langle PS |
\bar{\psi}(0) \, \gamma^+ \,  \varepsilon_T^{\nu\mu} \, S_{T \nu} \,
i\partial_{T \mu} \, \psi(y^-) | PS \rangle = T_F(x,x) \,,
\end{equation}
as well as
\begin{eqnarray} \label{eq:lc_2}
\lefteqn{\int \frac{dy^-dy_1^-}{4 \pi }P^+ e^{ixP^+ y^-}
e^{i(x-x_1)P^+y_1^-}\langle PS | \bar{\psi}(0) \, \gamma^{+} \,
\varepsilon_T^{\nu\mu} \, S_{T \nu} \, gA_{T \mu}(y_1^-) \,
\psi(y^-) | PS \rangle}
\nonumber \\
&& = \frac{i}{x-x_1+i\epsilon}\int \frac{dy^-dy_1^-}{4 \pi} \, 
e^{ixP^+y^-} e^{i(x-x_1)P^+y_1^-}\langle PS | \bar{\psi}(0) \, \gamma^{+} \,
\varepsilon_T^{\nu\mu} \, S_{T \nu} \, g F_{\phantom{+}\mu}^{+}(y_1^-) \, 
\psi(y^-) | PS \rangle \,.
\end{eqnarray}
One has to organize the contributions associated with $\bar{\psi} \partial_T \psi$ and 
$\bar{\psi} A_T \psi$ in a different way when using different boundary 
conditions~\cite{Zhou:2009jm,Zhou:2008mz}.
Though the final result is of course independent of the boundary condition, the 
calculation of the hard part associated with $\bar{\psi} A_T \psi$ in Drell-Yan is 
much more involved for both the advanced and the anti-symmetric boundary condition.
For a general discussion about these issues and some more technical details we refer 
the interested reader to a forthcoming paper~\cite{boundary}.

For the chiral-even contribution, the generalized factorization formula takes the form
\begin{eqnarray}
\frac{d \sigma (S_T)}{d Q^2 d\Omega}  & \propto &  
\frac{\alpha_{em}^2}{12 Q^2} \sum_{q} e_q^2
\int dx \, dx' \, T_F^q(x,x) \, f_1^{\bar{q}}(x') \, \varepsilon_T^{\rho \sigma} S_{T \rho}
\nonumber \\
& & \mbox{} \times 
\bigg[ \frac{\partial}{\partial k_T^\sigma} \Big(
H^{\mu \nu}_{Born}(xp+k_T,x'\bar{p}) \, V_{3,\mu \nu}^{CS} \, \sin 2 \theta \cos \phi
\nonumber \\
& & \hspace{1.3cm} +
H^{\mu \nu}_{Born}(xp+k_T,x'\bar{p}) \, V_{8,\mu \nu}^{CS} \, \sin 2 \theta \sin \phi
 \Big)_{k_T=0}
\nonumber \\ 
& & \hspace{0.5cm}
+ \frac{1}{\pi} \int dx_1 \, \frac{i}{x-x_1+i\epsilon} \,
H^{\mu \nu}_\sigma(xp,x_1p,x'\bar{p}) \, V_{8,\mu \nu}^{CM} \, \sin 2 \theta \sin \phi_S
\bigg] \,,
\end{eqnarray}
where, in the end, only the $k_T$-expansion of the hadronic tensor contracted with the 
tensor $V_{3,\mu\nu}^{CS}$ contributes to the asymmetry, while the corresponding 
expression associated with $V_{8,\mu\nu}^{CS}$ vanishes due to parity conservation.
This point is exactly reversed in the case of the chiral-odd part related with 
$T_F^{(\sigma)}$.
Note that the required imaginary part in the hard term coupled with the operator 
$\bar{\psi} A_T \psi$ can arise from the (artificial) pole $1/(x-x_1+ i \epsilon)$, 
which is generated by partial integration in Eq.~(\ref{eq:lc_2}).
Moreover, the diagrams with a special propagator contribute to the hard parts
resulting from both the $k_T$-expansion and the gluon-exchange.
However, these two contributions cancel each other.

The perturbative calculation is rather straightforward.
The final result for $A_N$ of the calculation in the light-cone gauge exactly matches 
with the final result~(\ref{eq:final}) we found in covariant gauge.

%
% 4. Section: Summary
% ==================
%
\section{Summary }
In summary, we recalculated the transverse single spin asymmetry $A_N$ in the 
angular distribution of a Drell-Yan dilepton pair by using twist-3 collinear 
factorization.
Compared to previous work on this topic, we payed particular attention to the 
$k_T$-dependence of the lepton tensor when making the collinear expansion.
Our final result for $A_N$ in Eq.~(\ref{eq:final}) differs from all the previous
results given in the literature.
For instance, we find an asymmetry which is just half of what was obtained in
Refs.~\cite{Boer:1997bw,Boer:1999si,Boer:2001tx}.

We made various checks in order to gain further confidence in our calculation.
First, we verified QCD gauge invariance by performing the calculation both in
covariant gauge and in the light-cone gauge.
Second, we tested the electromagnetic gauge invariance by recalculating the 
asymmetry in two specific QED light-cone gauges.
Third, we computed the NLO real emission corrections in the leading-log 
approximation.
In a certain sense, this calculation is more straightforward than the lowest
order treatment, since the $k_T$-flow can go through the unobserved parton line.
The outcome of this study fully supports our result for $A_N$ presented in the
present work.
A complete NLO analysis will be presented elsewhere.

It is important to notice that measuring the sign of $A_N$ can be considered to 
be equally important as checking the predicted sign reversal of the Sivers effect
in Drell-Yan~\cite{Collins:2002kn}.
In either case the physics of initial state gluon interactions would be tested.

The formalism developed in this paper can be extended in order to study similar 
observables which represent a correlation between the transverse spin and the 
relative transverse momentum of final state particles.
For instance, transverse SSAs for dihadron production in semi-inclusive DIS can,
in principle, be treated along the same lines.
We plan to address this point in a future work.
\\[0.3cm]
{\bf Acknowledgements}: We thank F. Yuan and D. Boer for helpful discussion.
This work is supported by the NSF under Grant No. PHY-0855501.

%%%%%%%%%%%%%%%%%%%%%%%%%%%%%%%%%%%%%%%%%%%%%%%%%%%%%%%%%%%%%%%%%%%%%%%%%%%%%%%%%%%%


\begin{thebibliography}{99}

%\cite{D'Alesio:2007jt}
\bibitem{D'Alesio:2007jt}
  U.~D'Alesio and F.~Murgia,
  %``Azimuthal and Single Spin Asymmetries in Hard Scattering Processes,''
  Prog.\ Part.\ Nucl.\ Phys.\  {\bf 61}, 394 (2008)
  [arXiv:0712.4328 [hep-ph]].
  %%CITATION = PPNPD,61,394;%%

%\cite{Burkardt:2008jw}
\bibitem{Burkardt:2008jw}
  M.~Burkardt, A.~Miller and W.~D.~Nowak,
  %``Spin-polarized high-energy scattering of charged leptons on nucleons,''
  Rept.\ Prog.\ Phys.\  {\bf 73}, 016201 (2010)
  [arXiv:0812.2208 [hep-ph]].
  %%CITATION = RPPHA,73,016201;%%

%\cite{Barone:2010ef}
\bibitem{Barone:2010ef}
  V.~Barone, F.~Bradamante and A.~Martin,
  %``Transverse-Spin and Transverse-Momentum Effects in High-Energy Processes,''
  Prog.\ Part.\ Nucl.\ Phys.\  {\bf 65}, 267 (2010)
  [arXiv:1011.0909 [hep-ph]].
  %%CITATION = PPNPD,65,267;%%

%\cite{Rogers:2010dm}
\bibitem{Rogers:2010dm}
  T.~C.~Rogers and P.~J.~Mulders,
  %``No Generalized TMD-Factorization in the Hadro-Production of High Transverse
  %Momentum Hadrons,''
  Phys.\ Rev.\  D {\bf 81}, 094006 (2010)
  [arXiv:1001.2977 [hep-ph]].
  %%CITATION = PHRVA,D81,094006;%%

%\cite{Efremov:1981sh}
\bibitem{Efremov:1981sh}
  A.~V.~Efremov and O.~V.~Teryaev,
  %``On Spin Effects In Quantum Chromodynamics,''
  Sov.\ J.\ Nucl.\ Phys.\  {\bf 36}, 140 (1982)
  [Yad.\ Fiz.\  {\bf 36}, 242 (1982)].
  %%CITATION = YAFIA,36,242;%%

%\cite{Efremov:1984ip}
\bibitem{Efremov:1984ip}
  A.~V.~Efremov and O.~V.~Teryaev,
  %``QCD Asymmetry And Polarized Hadron Structure Functions,''
  Phys.\ Lett.\  B {\bf 150}, 383 (1985).
  %%CITATION = PHLTA,B150,383;%%

%\cite{Qiu:1991pp}
\bibitem{Qiu:1991pp}
  J.~w.~Qiu and G.~Sterman,
  %``Single Transverse Spin Asymmetries,''
  Phys.\ Rev.\ Lett.\  {\bf 67}, 2264 (1991);
  Nucl.\ Phys.\  B {\bf 378}, 52 (1992);
  Phys.\ Rev.\  D {\bf 59}, 014004 (1999) [arXiv:hep-ph/9806356].
  %%CITATION = PRLTA,67,2264;%%

%\cite{Ellis:1982wd}
\bibitem{Ellis:1982wd}
  R.~K.~Ellis, W.~Furmanski and R.~Petronzio,
  %``Power Corrections To The Parton Model In QCD,''
  Nucl.\ Phys.\  B {\bf 207}, 1 (1982);
  Nucl.\ Phys.\  B {\bf 212}, 29 (1983).
  %%CITATION = NUPHA,B207,1;%%

%\cite{Jaffe:1991kp}
\bibitem{Jaffe:1991kp}
  R.~L.~Jaffe and X.~D.~Ji,
  %``Chiral odd parton distributions and polarized Drell-Yan,''
  Phys.\ Rev.\ Lett.\  {\bf 67}, 552 (1991);
  %%CITATION = PRLTA,67,552;%%
  %\cite{Jaffe:1991ra}
%\bibitem{Jaffe:1991ra}
  %``Chiral Odd Parton Distributions And Drell-Yan Processes,''
  Nucl.\ Phys.\  B {\bf 375}, 527 (1992).
  %%CITATION = NUPHA,B375,527;%%

%\cite{Kouvaris:2006zy}
\bibitem{Kouvaris:2006zy}
  C.~Kouvaris, J.~W.~Qiu, W.~Vogelsang and F.~Yuan,
  %``Single transverse-spin asymmetry in high transverse momentum pion
  %production in p p collisions,''
  Phys.\ Rev.\  D {\bf 74}, 114013 (2006)
  [arXiv:hep-ph/0609238].
  %%CITATION = PHRVA,D74,114013;%%

%\cite{Eguchi:2006mc}
\bibitem{Eguchi:2006mc}
  H.~Eguchi, Y.~Koike and K.~Tanaka,
  %``Twist-3 formalism for single transverse spin asymmetry reexamined:
  %Semi-inclusive deep inelastic scattering,''
  Nucl.\ Phys.\  B {\bf 763}, 198 (2007)
  [arXiv:hep-ph/0610314].
  %%CITATION = NUPHA,B763,198;%%

%\cite{Zhou:2009jm}
\bibitem{Zhou:2009jm}
  J.~Zhou, F.~Yuan and Z.~T.~Liang,
  %``Transverse momentum dependent quark distributions and polarized Drell-Yan
  %processes,''
  Phys.\ Rev.\  D {\bf 81}, 054008 (2010)
  [arXiv:0909.2238 [hep-ph]].
  %%CITATION = PHRVA,D81,054008;%%

%\cite{Hammon:1996pw}
\bibitem{Hammon:1996pw}
  N.~Hammon, O.~Teryaev and A.~Sch\"afer,
  %``Single spin asymmetry for the Drell-Yan process,''
  Phys.\ Lett.\  B {\bf 390}, 409 (1997)
  [arXiv:hep-ph/9611359].
  %%CITATION = PHLTA,B390,409;%%

%\cite{Boer:1997bw}
\bibitem{Boer:1997bw}
  D.~Boer, P.~J.~Mulders and O.~V.~Teryaev,
  %``Single spin asymmetries from a gluonic background in the Drell-Yan
  %process,''
  Phys.\ Rev.\  D {\bf 57}, 3057 (1998)
  [arXiv:hep-ph/9710223].
  %%CITATION = PHRVA,D57,3057;%%

%\cite{Boer:1999si}
\bibitem{Boer:1999si}
  D.~Boer and P.~J.~Mulders,
  %``Color gauge invariance in the Drell-Yan process,''
  Nucl.\ Phys.\  B {\bf 569}, 505 (2000)
  [arXiv:hep-ph/9906223].
  %%CITATION = NUPHA,B569,505;%%

%\cite{Boer:2001tx}
\bibitem{Boer:2001tx}
  D.~Boer and J.~w.~Qiu,
  %``Single transverse-spin asymmetry in Drell-Yan lepton angular
  %distribution,''
  Phys.\ Rev.\  D {\bf 65}, 034008 (2002)
  [arXiv:hep-ph/0108179].
  %%CITATION = PHRVA,D65,034008;%%

%\cite{Ma:2003ut}
\bibitem{Ma:2003ut}
  J.~P.~Ma and Q.~Wang,
  %``On unique predictions for single spin azimuthal asymmetry,''
  Eur.\ Phys.\ J.\  C {\bf 37}, 293 (2004)
  [arXiv:hep-ph/0310245].
  %%CITATION = EPHJA,C37,293;%%

%\cite{Anikin:2010wz}
\bibitem{Anikin:2010wz}
  I.~V.~Anikin and O.~V.~Teryaev,
  %``Gauge invariance, causality and gluonic poles,''
  Phys.\ Lett.\  B {\bf 690}, 519 (2010)
  [arXiv:1003.1482 [hep-ph]].
  %%CITATION = PHLTA,B690,519;%%

%\cite{Sivers:1989cc}
\bibitem{Sivers:1989cc}
  D.~W.~Sivers,
  %``Single Spin Production Asymmetries From The Hard Scattering Of Point-Like
  %Constituents,''
  Phys.\ Rev.\  D {\bf 41}, 83 (1990);
  Phys.\ Rev.\  D {\bf 43}, 261 (1991).
  %%CITATION = PHRVA,D41,83;%%

%\cite{Meng:1995yn}
\bibitem{Meng:1995yn}
  R.~Meng, F.~I.~Olness and D.~E.~Soper,
  %``Semi-Inclusive Deeply Inelastic Scattering at Small q_T,''
  Phys.\ Rev.\  D {\bf 54}, 1919 (1996)
  [arXiv:hep-ph/9511311].
  %%CITATION = PHRVA,D54,1919;%%

%\cite{Boer:2006eq}
\bibitem{Boer:2006eq}
  D.~Boer and W.~Vogelsang,
  %``Drell-Yan lepton angular distribution at small transverse momentum,''
  Phys.\ Rev.\  D {\bf 74}, 014004 (2006)
  [arXiv:hep-ph/0604177].
  %%CITATION = PHRVA,D74,014004;%%

%\cite{Berger:2007si}
\bibitem{Berger:2007si}
  E.~L.~Berger, J.~W.~Qiu and R.~A.~Rodriguez-Pedraza,
  %``Angular distribution of leptons from the decay of massive vector bosons,''
  Phys.\ Lett.\  B {\bf 656}, 74 (2007) [arXiv:0707.3150 [hep-ph]];
  Phys.\ Rev.\  D {\bf 76}, 074006 (2007) [arXiv:0708.0578 [hep-ph]].
  %%CITATION = PHLTA,B656,74;%%

%\cite{Arnold:2008kf}
\bibitem{Arnold:2008kf}
  S.~Arnold, A.~Metz and M.~Schlegel,
  %``Dilepton production from polarized hadron hadron collisions,''
  Phys.\ Rev.\  D {\bf 79}, 034005 (2009)
  [arXiv:0809.2262 [hep-ph]].
  %%CITATION = PHRVA,D79,034005;%%

%\cite{Collins:1977iv}
\bibitem{Collins:1977iv}
  J.~C.~Collins and D.~E.~Soper,
  %``Angular Distribution Of Dileptons In High-Energy Hadron Collisions,''
  Phys.\ Rev.\  D {\bf 16}, 2219 (1977).
  %%CITATION = PHRVA,D16,2219;%%

%\cite{Gottfried:1964nx}
\bibitem{Gottfried:1964nx}
  K.~Gottfried and J.~D.~Jackson,
  %``On the Connection between production mechanism and decay of resonances at
  %high-energies,''
  Nuovo Cim.\  {\bf 33}, 309 (1964).
  %%CITATION = NUCIA,33,309;%%

%\cite{Boer:1997nt}
\bibitem{Boer:1997nt}
  D.~Boer and P.~J.~Mulders,
  %``Time-reversal odd distribution functions in leptoproduction,''
  Phys.\ Rev.\  D {\bf 57}, 5780 (1998)
  [arXiv:hep-ph/9711485].
  %%CITATION = PHRVA,D57,5780;%%

%\cite{Boer:2003cm}
\bibitem{Boer:2003cm}
  D.~Boer, P.~J.~Mulders and F.~Pijlman,
  %``Universality of T-odd effects in single spin and azimuthal asymmetries,''
  Nucl.\ Phys.\  B {\bf 667}, 201 (2003)
  [arXiv:hep-ph/0303034].
  %%CITATION = NUPHA,B667,201;%%

%\cite{Qiu:1988dn}
\bibitem{Qiu:1988dn}
  J.~W.~Qiu,
  %``TWIST FOUR CONTRIBUTIONS TO THE PARTON STRUCTURE FUNCTIONS,''
  Phys.\ Rev.\  D {\bf 42}, 30 (1990).
  %%CITATION = PHRVA,D42,30;%%

%\cite{Zhou:2008mz}
\bibitem{Zhou:2008mz}
  J.~Zhou, F.~Yuan and Z.~T.~Liang,
  %``QCD Evolution of the Transverse Momentum Dependent Correlations,''
  Phys.\ Rev.\  D {\bf 79}, 114022 (2009)
  [arXiv:0812.4484 [hep-ph]].
  %%CITATION = PHRVA,D79,114022;%%

%\cite{Zhou:2009rp}
\bibitem{Zhou:2009rp}
  J.~Zhou, F.~Yuan and Z.~T.~Liang,
  %``Drell-Yan Lepton Pair Azimuthal Asymmetry in Hadronic Processes,''
  Phys.\ Lett.\  B {\bf 678}, 264 (2009)
  [arXiv:0901.3601 [hep-ph]].
  %%CITATION = PHLTA,B678,264;%%

\bibitem{boundary}
  J.~Zhou, F.~Yuan and Z.~T.~Liang, to be published.

%\cite{Collins:2002kn}
\bibitem{Collins:2002kn}
  J.~C.~Collins,
  %``Leading-twist Single-transverse-spin asymmetries: Drell-Yan and
  %Deep-Inelastic Scattering,''
  Phys.\ Lett.\  B {\bf 536}, 43 (2002)
  [arXiv:hep-ph/0204004].
  %%CITATION = PHLTA,B536,43;%%

\end{thebibliography}
\end {document}